\documentclass[pre,twocolumn,showpacs,preprintnumbers,amsmath,amssymb]{revtex4}
\usepackage[dvips]{graphicx}
\usepackage[dvips]{color}
\usepackage{amsmath}
\usepackage{amssymb}
\usepackage{dcolumn}
\usepackage{bm}
\allowdisplaybreaks

\begin{document}

\title{
Tumbling motion of a single chain in shear flow: a crossover from Brownian to non-Brownian behavior
}
\author{Hideki Kobayashi}
\email{hidekb@cheme.kyoto-u.ac.jp}
\author{Ryoichi Yamamoto}
\email{ryoichi@cheme.kyoto-u.ac.jp}
\affiliation{Department of Chemical Engineering, Kyoto University, Kyoto
615-8510, Japan}
\affiliation{CREST, Japan Science and Technology Agency , Kawaguchi 332-0012, Japan}

\date{\today}

\begin{abstract}
We present the numerical results for the dynamics of a single chain in
 steady shear flow. The chain is represented by a bead-spring model, and
 the smoothed profile method is used to accurately account for the
 effects of thermal fluctuations and hydrodynamic interactions acting on
 beads due to host fluids. It was observed that the chain undergoes
 tumbling motions and that its dimensionless frequency $F = 6\pi\eta
 \sigma^{3}\nu/k_{\rm B}T$ depends only on the Peclet number $Pe$ with a
 power law $F \propto Pe^{\alpha}$, where $k_{\rm B}$ is the Boltzmann
 constant, $T$ is the temperature, and $\sigma$ is the diameter of the
 beads. The exponent $\alpha$ clearly changes from $2/3$ to $1$ around
 the critical Peclet number, $Pe_{\rm c}$, indicating that the crossover
 reflects the competition of thermal fluctuation and shear flow. The
 presented numerical results agree well with our theoretical analysis
 based on Jeffrey's work.
\end{abstract}

\maketitle

\section{INTRODUCTION}

The dynamics of solid particles dispersed in host fluids is an important
problem in many different fields of science and engineering. The
macroscopic properties of such dispersions (the elastic modulus, viscosity,
and thermal and electric conductivities) greatly depend on the dynamics of
the particles in the host fluids. In equilibrium states, the dynamics of
small dispersed particles are strongly affected by the thermal
fluctuations of their host fluids. When flow is imposed, the dynamics are
also affected by the flow of the host fluids. Because it is difficult to
experimentally analyze these complex particle dynamics, which are
coupled both to thermal fluctuations and to fluid flow, numerical
simulations are particularly important for understanding the properties
of particle dispersions in detail \cite{kamata}.

For a single Brownian chain fluctuating in the shear flow of a Newtonian
fluid, it has been suggested that the tumbling frequency $\nu$ is
proportional to the shear rate $\dot{\gamma}^{2/3}$ \cite{
dna_exp1, dna_exp2, dna_exp3, flexible_sim1, flexible_sim2,
flexible_theo1, flexible_theo2, flexible_theo3}. This has been
experimentally confirmed by Schroeder, Teixeira, Shaqfeh, and Chu
\cite{dna_exp1} for the dynamics of individual DNA molecules in
a linear shear flow. Primarily, Smith, Babcock, and Chu \cite{chu_science}
measured the power spectral density (PSD) and the probability
distribution function (PDF) of the extension length of each DNA molecule
for various Weissenberg numbers $Wi=\dot{\gamma}\tau$, where
$\dot{\gamma}$ is the shear rate and $\tau$ is the relaxation time of
the chain orientation. The PSD of polymer extension exhibits no peaks.
In subsequent experiments \cite{dna_exp1,
dna_exp2}, however, the focus has been put on the PSD of the orientation
angle $\varphi$, where $\varphi = 0$ when the DNA molecule lies
perfectly in the flow direction. These experimental results support a
simple power law, $\nu \tau \propto Wi^{2/3}$, where $\nu$ is the peak
frequency of the PSD. The relaxation time $\tau$ is considered to be a
constant if the temperature is constant. This leads to
$\nu\propto\dot{\gamma}^{2/3}$. Similar results have also been obtained
in other experiments \cite{dna_exp2, dna_exp3}, numerical simulations
\cite{flexible_sim1, flexible_sim2}, and theoretical analyses
\cite{flexible_theo1, flexible_theo2, flexible_theo3}. Although the DNA
molecules mentioned above can be considered flexible chains, a quite
similar power law was obtained using numerical simulations for a Brownian
linear rigid rod as well \cite{dna_exp1}.

For a single non-Brownian (thermally non-fluctuating) flexible chain in
shear flow, the tumbling frequency $\nu$ is expected to be
proportional to the shear rate $\dot{\gamma}$
\cite{deformation_fibre_sim}. A single non-Brownian rigid rod with a
finite aspect ratio is known to exhibit a cyclic tumbling motion in
shear flow, as described by 
Jeffrey's equation, $\nu\propto\dot{\gamma}$ \cite{jeffery, breth, free_rotation}.

 From the above experimental findings, one would expect to observe a
crossover from Brownian ($\nu\propto\dot{\gamma}^{2/3}$) to non-Brownian
($\nu\propto\dot{\gamma}$) behavior with increasing shear rate; however,
such a clear crossover has not yet been reported. This crossover has not
yet been successfully predicted by previous numerical
\cite{flexible_sim1, flexible_sim2} and theoretical
\cite{flexible_theo1, flexible_theo2, flexible_theo3} studies where a
dispersed chain is treated as an end-to-end vector, namely, as an
infinitely thin line. Therefore, the rotational motion of the chain
cannot be sustained across $\varphi=0$ without thermal
fluctuations. When a thermal fluctuation exists, the orientation of the
thin line can fluctuate around $\varphi=0$.\ This leads to tumbling
motions even for a thin line; however, the frequency of the tumbling motion
$\nu$ is always proportional to $\dot{\gamma}^{2/3}$, regardless of the
shear rate \cite{flexible_theo3}.

We thus aim to analyze this crossover by use of a direct numerical
simulation (DNS) approach. In the present study, we simulated the
tumbling motion of a chain using a smoothed profile method (SPM) that
accurately takes into account thermal fluctuations and hydrodynamic
interactions \cite{spm, spm2, spm_fl, spm_fl2}. The chain is represented by a
bead-spring model, wherein each bead is modeled as a spherical object
with a finite radius $a$ and undergoes free rotation. Rigid rods or
flexible chains are represented with or without a constraint force on
bond bending. Apart from the previous numerical models, the presented
bead-spring model naturally takes into account the finite thickness of
the experimentally used chains or rods. A theoretical analysis has also
been developed to understand the mechanisms underlying the crossover.

\section{METHODS}

\subsection{MODEL}

We solve the dynamics of a single chain in a Newtonian solvent using SPM
\cite{spm, spm2, spm_fl, spm_fl2}.
In this method, boundaries between solid particles and solvents are
replaced with a continuous interface by assuming a smoothed
profile. This enables us to calculate hydrodynamic interactions both
efficiently and accurately, without neglecting many-body interactions.
The equation governing a solvent with a density
$\rho_{\rm f}$ and a shear viscosity $\eta$ is a modified Navier-Stokes
equation,

\begin{eqnarray}
 \rho_{\rm f}  \left\{ \frac{\partial \vec{v}}{\partial t}+ 
   (\vec{v}\cdot\vec{\bigtriangledown})\vec{v} \right\} =
 -\vec{\bigtriangledown}p + \eta \vec{\bigtriangledown}^2\vec{v} +
 \rho_{\rm f}\phi\vec{f_{\rm p}} + \vec{f}_{\rm shear}
\end{eqnarray}
with the incompressible condition
$\vec{\bigtriangledown}\cdot\vec{v}=0$, where $\vec{v}(\vec{r},t)$ and
$p(\vec{r},t)$ are the velocity and pressure fields of the solvent,
respectively. A smoothed profile function $0\leq\phi(\vec{r},t)\leq1$
distinguishes between the fluid and particle domains, yielding $\phi=1$
in the particle domain and $\phi=0$ in the fluid domain. These domains
are separated by thin interstitial regions, the thicknesses of which are
characterized by $\xi$. The body force $\phi\vec{f_{\rm p}}$ is
introduced to ensure the rigidity of the particles and the appropriate
non-slip boundary condition at the fluid/particle interface. The
mathematical expressions for $\phi$ and $\phi\vec{f_{\rm p}}$ are
detailed in previous papers \cite{spm,spm2}. The external force
$\vec{f}_{\rm shear}$ is introduced to maintain a linear shear
\cite{shear}, expressed by

\begin{eqnarray}
 v_x=\left\{
      \begin{array}{lll}
-\dot{\gamma}y& (0<y<L_y/4),\\
-\dot{\gamma}\left(-y+\dfrac{L_y}{2}\right)& (L_y/4<y<3L_y/4),\\
-\dot{\gamma}\left(y-L_y\right)& (3L_y/4<y<L_y),\\
      \end{array}
     \right.
\end{eqnarray}
where $\dot{\gamma}$ is the shear rate and $L_y$ is the system size in
the $y$-direction.

In the present study, the chain is represented as either a rigid rod or
a flexible chain. We use a bead-spring model consisting of $N$ beads
in a single chain. The bead size is sufficient to fit several mesh
units. Therefore, it is necessary to consider the torque exerted on the
bead. The motion of the $i$th bead is governed by the following Newton's
and Euler's equations of motion with stochastic forces:

\begin{eqnarray}
M_i\frac{d}{dt}\vec{V}_i=\vec{F}_i^{\rm H}+\vec{F}_i^{\rm P}+\vec{F}_i^C+\vec{G}_i^V,\;\;\;
  \frac{d}{dt}\vec{R}_i=\vec{V}_i,
\end{eqnarray}

\begin{eqnarray}
 \vec{I}_i\cdot\frac{d}{dt}\vec{\Omega}_i=\vec{N}_i^{\rm H}+\vec{G}_i^{\Omega},
\end{eqnarray}
where $\vec{R}_i$, $\vec{V}_i$ and $\vec{\Omega}_i$ are the position, translational velocity, and rotational velocity of the beads, respectively. $M_i$ and $\vec{I_i}$ are the mass and moment of inertia, and $\vec{F}_i^{\rm H}$ and $\vec{N}_i^{\rm H}$ are the hydrodynamic force and torque exerted by the solvent on the beads, respectively \cite{spm,spm2}. $\vec{G}_i^{\rm V}$ and $\vec{G}_i^{\Omega}$ are the random force and torque, respectively, due to thermal fluctuations. The temperature of the system is defined such that the long-time diffusive motion of dispersed particles reproduces correct 
behavior \cite{spm_fl,spm_fl2}.

$\vec{F}_i^{\rm P}$ represents the potential force due to direct
inter-bead interactions, such as Coulombic and Lennard-Jones
potentials. We use a bead-spring model as a model of a polymeric chain
with a truncated Lennard-Jones potential and a finitely extensible
nonlinear elastic (FENE) potential. The truncated Lennard-Jones 
interaction is expressed in terms of $U_{\rm LJ}$:

\begin{eqnarray}
   U_{\rm LJ}(r_{ij})=\left\{
		   \begin{array}{lll}
		    4\epsilon \left\{
			       \left( \dfrac{\sigma}{r_{ij}}\right)^{12} -
			       \left( \dfrac{\sigma}{r_{ij}}\right)^{6}\right\}
		    + \epsilon &(r_{ij}<2^{\frac{1}{6}}\sigma), \\
		    0 &(r_{ij}>2^{\frac{1}{6}}\sigma),
		   \end{array}
			\right.
\end{eqnarray}
where $r_{ij}=|\vec{R}_i - \vec{R}_j|$. The parameter $\epsilon$
characterizes the strength of the interactions, and $\sigma$ represents
the diameter of the beads. Consecutive beads on a chain are connected by
a FENE potential of the form

\begin{eqnarray}
U_{\rm FENE}(r)=-\frac{1}{2}k_{\rm c}R_0^2\ln\left\{1-\left(\frac{r}{R_0}\right)^2\right\},
\end{eqnarray}
where $r=|\vec{R}_{i+1} - \vec{R}_i|$, $k_c=30\epsilon/\sigma^2$, and
$R_0=1.5\sigma$.
$\vec{F}_i^{\rm C}$ is the constraint force acting on the $i$th bead due
to the bond-angle constraints that cause the chain to form a straight
line, and it is used only for the rigid rod case. This is given by 

\begin{eqnarray}
   \vec{F_i^{\rm C}}=\frac{\partial}{\partial\vec{R_i}}(\displaystyle
    \sum^N_{\alpha=3} \vec{\mu}_{\alpha}\cdot\vec{\Psi}_{\alpha}),
\end{eqnarray}

\begin{eqnarray}
   \vec{\Psi}_{\alpha}=(\alpha-2)\vec{R}_1-(\alpha-1)\vec{R}_2+\vec{R}_\alpha,
    \;\;\;\ (\alpha=3,\cdots,N),
    \label{constraint}
\end{eqnarray}
where $\vec{\Psi}_{\alpha}=0$ is the constraint condition to be
satisfied. $\vec{\mu}_{\alpha}$ is a Lagrange multiplier associated with
the intramolecular forces of the constraints chosen such that the
constraint condition $\vec{\Psi}_{\alpha}=0$ is satisfied at a time
$t+h$, where h is the time increment of a single simulation step.

\subsection{SIMULATION}

Numerical simulations have been performed in three dimensions with
periodic boundary conditions. The lattice spacing $\Delta$ is taken to
be the unit of length. The unit of time is given by $\rho_{\rm
f}\Delta^2/\eta$, where $\eta=1$ and $\rho_{\rm f}=1$. The system size
is $L_x\times L_y\times L_z=32\times16\times64$. The other parameters
include: $\sigma=4$, $\xi=2$, $\epsilon=1$, $\eta=1$, $M_i=4\pi a^3/3$,
$N=5$ and $h=0.067$.

In the presented simulations, the Navier-Stokes equation is discretized with
a de-aliased Fourier spectral scheme in space and with a second-order
Runge-Kutta scheme in time. To follow bead motions, the
position, velocity and angular velocity of the beads are integrated with
the Adams-Bashforth scheme.

At $t=0$, the chain aligns along the $x$-axis, which is the shear
direction. The run-time of our simulations is about
$3520/\dot{\gamma}$. The range of $k_{\rm B}T$ is
$5.0\times10^{-4}<k_{\rm B}T<1.0$, and that of $\dot{\gamma}$ is
$1.0\times10^{-3}<\dot{\gamma}<4.0\times10^{-2}$.

From the symmetry of the system, to analyze the tumbling motion of a
chain, we only have to consider the projected tumbling motion on the
$x$-$y$ plane. We introduce the chain orientation angle $\varphi$, which
is the angle between the $x$-axis and the projected end-to-end vector on
the $x$-$y$ plane.

Evidence of periodic tumbling motion should appear in the PSD per unit time \cite{dna_exp1, dna_exp2}. Therefore, to
investigate the spectral properties of the orientation, we calculate the
PSD using a fast Fourier transform. The PSD is expressed as
\begin{eqnarray}
 PSD(\omega)= |\int \varphi(t)\exp(i\omega t)dt|. 
\end{eqnarray}

\section{RESULTS}

\begin{figure}[t]
 \begin{center}
  \includegraphics[width=1.0\hsize]{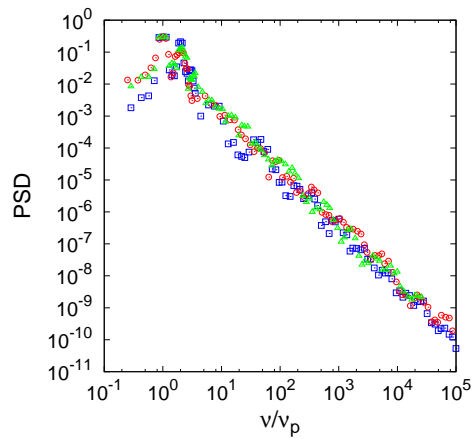}
%Fig.1
  \caption{(Color online)
  The behavior of PSD as a function of $\nu / \nu_p$ for rigid rods at
  $k_{\rm B}T=0.0005$ and $\dot{\gamma}=0.001$ (blue $\Box$), 
  $k_{\rm B}T=0.006$ and $\dot{\gamma}=0.002$ (red $\bigcirc$), and $k_{\rm B}T=0.01$
  and $\dot{\gamma}=0.008$ (green $\bigtriangleup$).
  }
  \label{fig_psd_f_rigid}
 \end{center}
\end{figure}

\begin{figure}[t]
 \begin{center}
  \includegraphics[width=1.0\hsize]{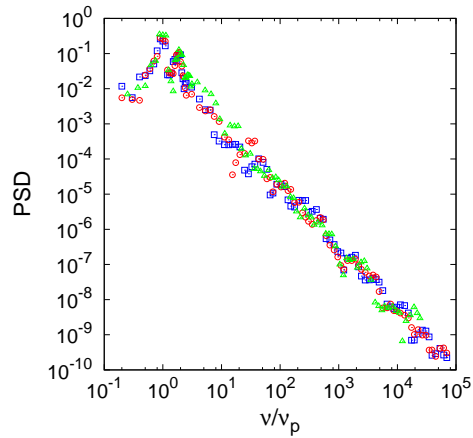}
%Fig.2
  \caption{(Color online)
  The behavior of PSD as a function of $\nu / \nu_p$ for flexible chains
  at $k_{\rm B}T=0.0005$ and $\dot{\gamma}=0.001$ (blue $\Box$), $k_{\rm
  B}T=0.006$ and $\dot{\gamma}=0.002$ (red $\bigcirc$), and $k_{\rm B}T=0.01$
  and $\dot{\gamma}=0.008$ (green $\bigtriangleup$).
  }
  \label{fig_psd_f_flexible}
 \end{center}
\end{figure}

In Figs. \ref{fig_psd_f_rigid} and \ref{fig_psd_f_flexible}, the PSDs of
the chain orientation angle $\varphi$ show a peak at a specific
frequency $\nu_p$, as shown in previous studies \cite{dna_exp1,
dna_exp2}. Furthermore, the PSD data obtained at different conditions of
shear rate and temperature tend to lie on a single master curve if a
normalized frequency $\nu / \nu_p$ is used. This is true for both rigid
rods and flexible chains. This result implies that the tumbling motion
of chains is fully characterized by $\nu_p$.

\begin{figure}[t]
 \begin{center}
  \includegraphics[width=1.0\hsize]{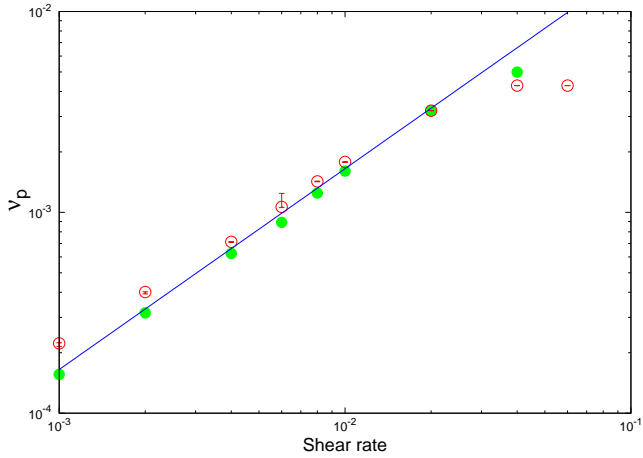}
%Fig.3
  \caption{(Color online)
  The behavior of $\nu_p$ as a function of $\dot{\gamma}$ at $k_{\rm
  B}T=0.0005$. Rigid rod (green closed circle), flexible chain (red open
  circle). The blue solid line corresponds to the law
  $\nu_p\propto\dot{\gamma}$. Error bars are
  derived from the half band width of PSD.
  }
  \label{fig_kbt00005}
 \end{center}
\end{figure}

\begin{figure}[t]
 \begin{center}
  \includegraphics[width=1.0\hsize]{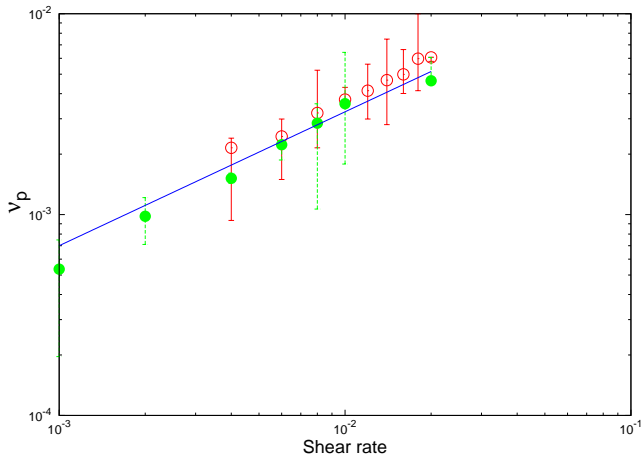}
%Fig.4
  \caption{(Color online)
  The behavior of $\nu_p$ as a function of $\dot{\gamma}$ at
  $k_{\rm B}T=1.00$.
  Rigid rod (green closed circle), flexible chain (red open circle). The
  blue solid
  line corresponds to the law $\nu_p\propto\dot{\gamma}^{2/3}$.
  Error bars are
  derived from the half band width of PSD.
  }
  \label{fig_kbt100}
 \end{center}
\end{figure}

In Fig. \ref{fig_kbt00005}, we find that $\nu_p$ follows the law $\nu_p
\propto \dot{\gamma}$ for $\dot{\gamma}<0.02$ at $k_{\rm
B}T=5.0\times10^{-4}$, for both the rigid rod and the flexible chain
cases. Meanwhile, Fig. \ref{fig_kbt100} shows $\nu_p \propto
\dot{\gamma}^{0.68}$ for $\dot{\gamma}<0.02$ at $k_{\rm B}T=1.0$ for
both.

Figure  \ref{fig_kbt00005} also shows that, for $\dot{\gamma}>0.02$, $\nu_{p}$ is lower
than the frequency expected from the law $\nu_p \propto
\dot{\gamma}$. P. Bagchi and S. Balachandar have reported that, at a
finite Reynolds number approximately equal to 1, the sphere rotation
frequency in a linear shear flow decreases at a much slower rate than
$\dot{\gamma}/2$ \cite{moderate_re}. The Reynolds number, $Re$, is given
by $Re=\rho_{f}\dot{\gamma}\sigma^{2}N/\eta$. In our paper, $Re$ is
equal to $1.6$ at $\dot{\gamma}=0.02$. We only consider $\nu_p$ in the
region of $\dot{\gamma}\leq0.02$, so the effects of finite Reynolds
numbers do not influence the results.

We arrange data sets using the Peclet number to consider the effect of
competition between shear and fluctuation. The Peclet number is the
dimensionless number that relates the rate of shear flow to the rate of
thermal fluctuation. In our work, the Peclet number, $Pe$, and
dimensionless frequency, $F$, are expressed as

\begin{eqnarray}
 Pe = \frac{6\pi\eta \sigma^{3}\dot{\gamma}}{k_{\rm B}T},
  \label{def_peclet}
\end{eqnarray}

\begin{eqnarray}
 F = \frac{6\pi\eta \sigma^{3}\nu_p}{k_{\rm B}T}.
  \label{def_freq}
\end{eqnarray}

\begin{figure}[t]
   \begin{center}
    \includegraphics[width=1.0\hsize]{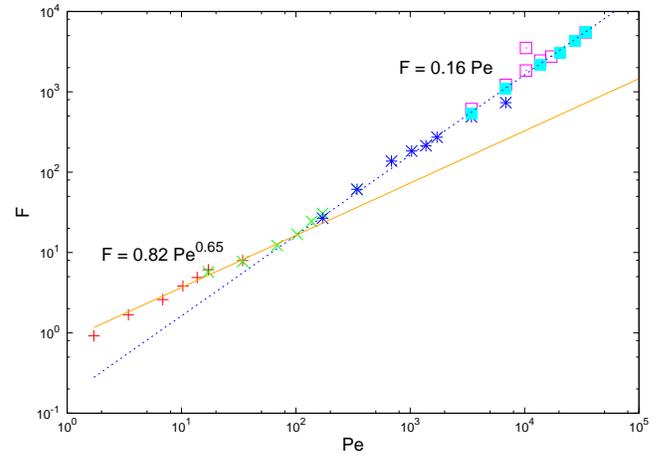}
%Fig.5
    \caption{(Color online)
    The behavior of $F$ as a function of $Pe$ for a rigid rod. The
    orange solid
    and blue dotted lines were calculated from a least-squares fit of
    the data points. 
    The orange solid line corresponds to $Pe^{0.65}$. The blue dotted line
    corresponds to $Pe$.
}
    \label{fig_pe_rigid}
   \end{center}
  \end{figure}

\begin{figure}[t]
   \begin{center}
    \includegraphics[width=1.0\hsize]{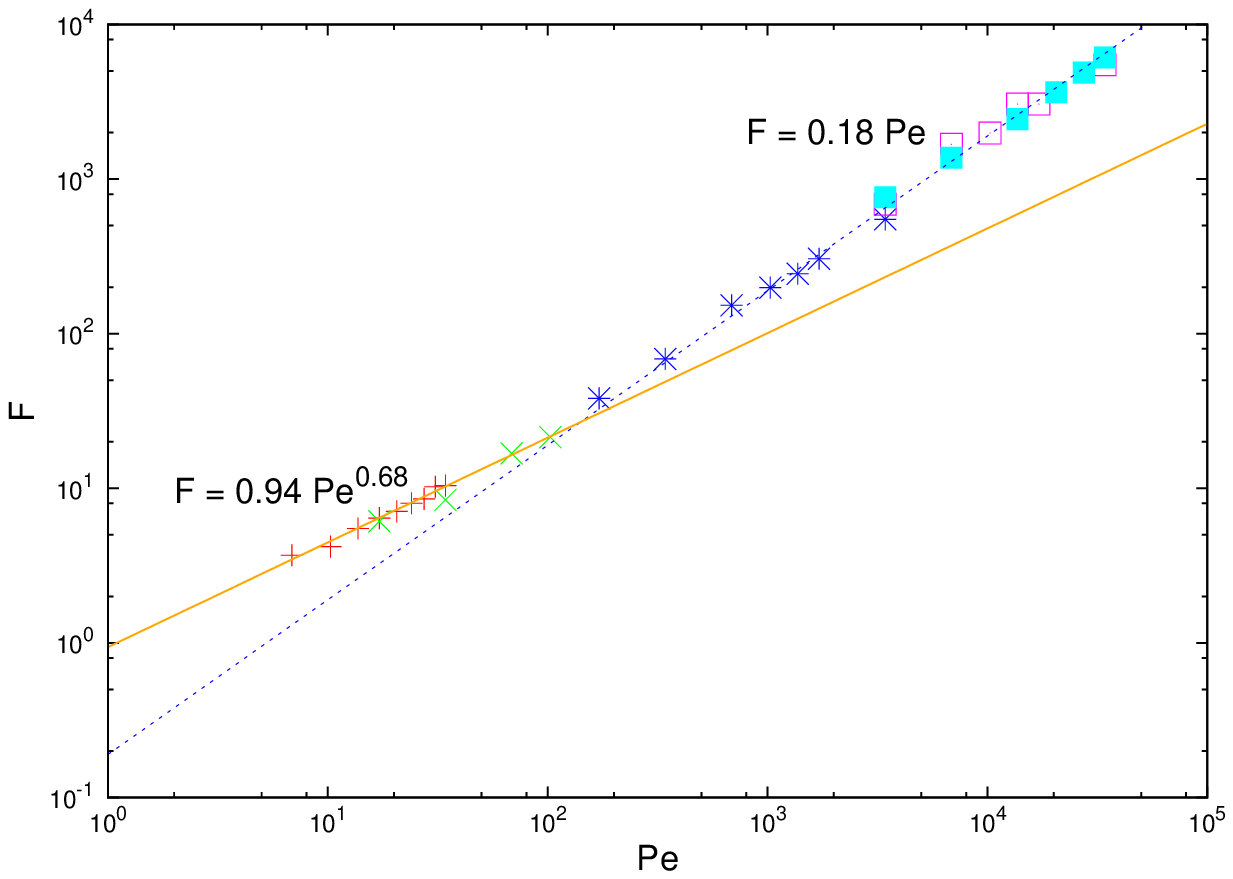}
%Fig.6
    \caption{(Color online)
    The behavior of $F$ as a function of $Pe$ for a flexible chain. The
    orange solid and blue dotted lines were calculated from a
    least-squares fit of the data 
    points. The orange solid line corresponds to $Pe^{0.68}$. The blue
    dotted line corresponds to $Pe$.
}
    \label{fig_pe_flexible}
   \end{center}
  \end{figure}

We plotted the behavior of $F$ as a function of $Pe$ for the rigid rod and
flexible chain cases in Figs. \ref{fig_pe_rigid} and
\ref{fig_pe_flexible}. $F$ was found to depend only on $Pe$ because
the data sets have the same value of $F$ with the same value of $Pe$,
even when the shear rates and temperatures are different. In the rigid rod
case, $F \propto Pe^{0.65}$ for $Pe<106$ and $F\propto Pe$
for $Pe>106$. In the flexible chain case, $F \propto
Pe^{0.68}$ for $Pe<156$ and $F\propto Pe$ for $Pe>156$. The behaviors of
$F$ for the rigid rod case and the flexible chain case are roughly
equal, although the values of the Peclet numbers are different at the
boundary where the exponent of $Pe$ changes. 

We define the critical value at which the exponent of $Pe$ drastically
changes from almost 2/3 to 1 as the {\it critical Peclet number},
$Pe_{\rm c}$. $F \propto Pe^{2/3}$ for $Pe<Pe_{\rm c}$; otherwise,
$F\propto Pe$. When fluctuations dominate the system, the exponent is
nearly equal to $2/3$. On the other hand, when shear flow dominates, the
exponent is exactly equal to $1$.

  \begin{figure}
   \begin{center}
    \includegraphics[width=1.0\hsize]{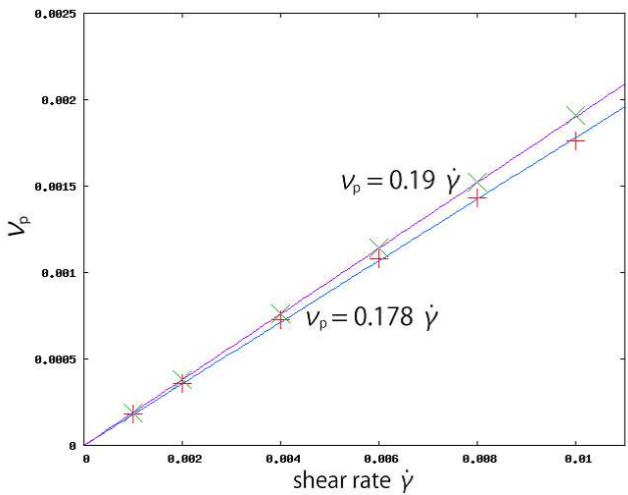}
%Fig.7 
    \caption{(Color online)
    The behavior of $\nu_p$ as a function of $\dot{\gamma}$ at
    $k_{\rm B}T=0$. Rigid rod (red $+$), flexible chain (green
    $\times$).
    The data sets
    for the rigid rods correspond to the law $\nu_p = 0.178
    \dot{\gamma}$. The data sets for the flexible chains correspond to
    the law $\nu_p = 0.190 \dot{\gamma}$.
}
    \label{fig_kbt0}
   \end{center}
  \end{figure}

In order to understand the behavior of $F$ in the limit of
$Pe\rightarrow\infty$, we examine the behavior of $\nu_p$ in the limit of
$k_{\rm B}T\rightarrow0$. As shown in Figs. \ref{fig_pe_rigid} and
\ref{fig_pe_flexible}, the behavior of $F$ in the region of $Pe_{\rm c} <
Pe$ is roughly equal to the behavior of $\nu_p$ in the limit of $k_{\rm
B}T\rightarrow0$. Therefore, we believe that the proportional relation
$F \propto Pe$ can be true across the entire region of $Pe_{\rm c} < Pe$.

\section{DISCUSSION}

\subsection{COMPARISON WITH OTHER RESULTS}

In our work, we calculate the tumbling motion of a single chain for
$0.0005<k_{\rm B}T<1.00$ and $0.001<\dot{\gamma}<0.04$. As reported in
Jeffrey's paper \cite{jeffery}, which treated non-Brownian particles,
$\nu_p$ follows the law $\nu_p\propto\dot{\gamma}$ at $k_{\rm B} T
=5.0\times10^{-4}$, as shown in Fig. \ref{fig_kbt00005}. In previous
papers \cite{flexible_theo1, flexible_theo2, flexible_sim1,
flexible_sim2} that treated Brownian particles, $\nu_p$ follows the law
$\nu_p\propto\dot{\gamma}^{2/3}$ at $k_{\rm B} T = 1.0$, as shown in
Fig. \ref{fig_kbt100}.

The dimensionless frequency, $F$, depends only on $Pe$. This $Pe$
dependence on $F$ can be described by a power law, $F \propto
Pe^{\alpha}$. The exponent $\alpha$ drastically changes from $2/3$ to
$1$ at $Pe_{\rm c}$; $\alpha$ equals $2/3$ for $Pe<Pe_{\rm c}$, while
$\alpha$ equals $1$ for $Pe_{\rm c}<Pe$. In the case of a rigid rod with
$N=5$, $Pe_{\rm c}\approx106$, and $Pe_{\rm c}\approx156$ in the
flexible chain case. When fluctuations dominate the system, $F$ follows
the law of $F \propto Pe^{2/3}$. On the other hand, when shear flow
dominates, $F$ follows the law of $F \propto Pe$.

Gerashchenko and Steinberg \cite{dna_exp3} claim that there
are two dynamical regimes of polymer motion at $Wi \gg 1$, depending on the polymer
extension $R$. When $R \ll R_{max}$, where $R_{max}$ is the
maximum polymer extension, the tumbling frequency $\nu$ is
constant and independent of $Wi$. On the other hand, when
$R \sim R_{max}$, $\nu$ is proportional to $Wi^{2/3}$. 
In Fig. 4, we did not observe $\nu_{p}$ to be independent of
$\dot{\gamma}$. We consider the chain length to be too short in our work.
The chain cannot keep the coil state along the shear direction for long times.
The chain is fully stretched at short notice and is always rotated,
although $Pe$ is small.

Previously, Szymczak and Cieplak \cite{proteins_shear} discussed
the conformational dynamics of a single long protein in shear flow and
found two characteristic tumbling frequencies, $\nu_1$ and
$\nu_2$. They showed that the higher frequency $\nu_1$ follows
the law of $\nu_1 \propto \dot{\gamma}$; however, the lower frequency
$\nu_2$ follows the law of $\nu_2 \propto \dot{\gamma}^{2/3}$. When the
protein is tightly packed, it essentially shows a spherical rotation in
shear flow. As a result, $\nu_1$ is proportional to $\dot{\gamma}$. The
lower frequency, $\nu_2$, corresponds to the stretching-collapse cycle;
hence, $\nu_2$ is proportional to $\dot{\gamma}^{2/3}$. Although our
results are similar to theirs, the phenomena in our system are
essentially different from those in their works because rigid rods
cannot fold.

Davoudi and Schumacher \cite{turbulent} analyzed the stretching of polymers in a
turbulent flow. It is known that the polymers undergo a coil-stretch
transition at $Wi\approx1/2$ in this system.
For $Wi<1/2$, polymers are in the coiled state, and
their size distribution is stationary.
In contrast, for $Wi>1/2$, the polymers are in the stretched state.
Their stretching carries on until their lengths reach the finite extensibility limit
or until turbulence stops the growth of the polymers.
They found the maximum Lyapunov exponent to be
$\lambda \sim \dot{\gamma}^{3/2}$. However, Chertkov et al. reported
$\lambda \sim \dot{\gamma}^{2/3}$ in their work, where $\lambda^{-1}$ is
expressed as the mean stretching time scale.
Davoudi and Schumacher claimed also that their study  could not
be compared with the analytic results of Chertkov et al.
They defined the shear time scale as $T_{s} = \dot{\gamma}^{-1}$ and the fluctuation time scale as
$T_{f} = D^{-1}$, where $D$ is the strength of Gaussian fluctuation.
Chertkov analyzed the polymer dynamics in the region of $T_{s} \ll T_{f}$.
Davoudi analyzed the polymer dynamics in the region of $\tau_{\eta}<T_{s}$, 
where $\tau_{\eta}$ is the Kolmogorov time.
Notably, Davoudi's work studies a different regime of polymer stretching
than the analytic model of Chertkov's work.
Chertkov et al. studied in the shear-dominated regime, whereas
Davoudi and Schumacher studied in the turbulence-dominated regime.

Our work analyzed polymer dynamics in the region of $T_{s} \ll T_{f}$.
We do not consider the effect of turbulent flow. We only considered the
region with particle Reynolds number $Re < 1.6$, so the effects of
finite Reynolds numbers do not influence the results.
In Davoudi's work, by contrast, the region of particle Reynolds
number in the stretched state is estimated to be $Re^{'} > 15$,
where $Re^{'}=vL/\nu$, $\nu$ is kinetic viscosity, $v$ is
the root-mean-square of the turbulent velocity fluctuation, and
$L$ is the mean length of polymer. Therefore, our study cannot be
compared with Davoudi's work.

\subsection{THEORETICAL ANALYSIS}

 \begin{figure}[t]
  \begin{center}
    \includegraphics[width=1.0\hsize]{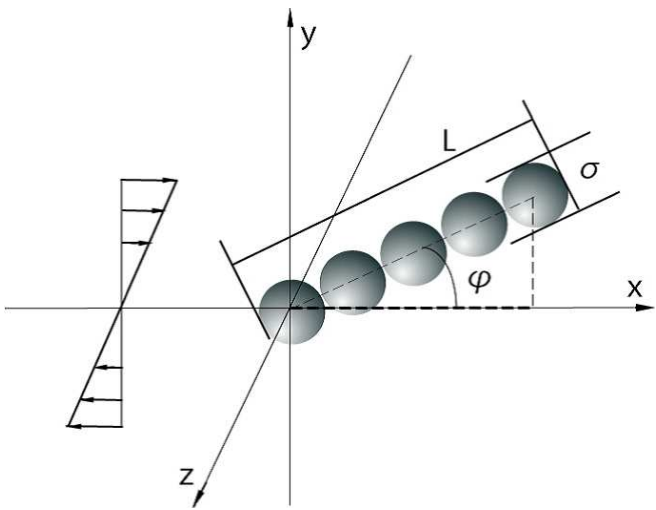}
%Fig.8
   \caption{The geometry of the present simulations.
   }
   \label{fig_bead_spring}
  \end{center}
 \end{figure}

The geometry of the chain in our paper is depicted in
Fig. \ref{fig_bead_spring}. In Jeffrey's work \cite{jeffery}, the angle
$\varphi$ of non-Brownian rigid rods with finite aspect ratios is
governed by the equation

\begin{eqnarray}
  \frac{d}{dt}\varphi=
   -\dot{\gamma}\frac{r^2-1}{r^2+1}\sin^2\varphi -
   \dot{\gamma}\frac{1}{r^2+1},
   \label{jeffrey_original}
 \end{eqnarray}
where the aspect ratio is $r=L/\sigma$ and $L$ is the length of the
chain.

In our work, we consider the equation that governs the angle $\varphi$
of a Brownian rigid rod with a finite aspect ratio. To consider
the diffusion of thermal fluctuation, we introduce white noise
$\zeta(t)$ into Eq. (\ref{jeffrey_original}). We can then write down the
following equation for a Brownian rigid rod:

\begin{eqnarray}
  \frac{d}{dt}\varphi=
   -\dot{\gamma}\frac{r^2-1}{r^2+1}\sin^2\varphi -
   \dot{\gamma}\frac{1}{r^2+1}
   + 2\sqrt{\frac{D_r}{\cos^2 \theta}}\zeta,
   \label{jeffrey_rod}
\end{eqnarray}

\begin{eqnarray}
  <\zeta(t)\zeta(t')>=\delta(t-t'),
\end{eqnarray}
where $D_r$ is the rotational diffusion constant. On the basis of the
shell model \cite{dr1, dr2}, the rotational diffusion constant $D_r$ for
a rigid rod is calculated as 

\begin{eqnarray}
  D_r=\frac{3(\ln r + d (r))k_{\rm B}T}{\pi \eta L^3},
\end{eqnarray}

\begin{eqnarray}
  d (r) = - 0.662 + \frac{0.917}{r} - \frac{0.05}{r^2}.
\end{eqnarray}

In the shell model mentioned above, 
the contour of the macromolecules of arbitrary shape is
represented by a shell composed of  many identical small beads.
The shell model can be adequately modeled by decreasing the size of the
beads.

In the case of $D_r\ll\dot{\gamma}$, the dynamics of the angle $\varphi$
become decoupled from the angle $\theta$ between the end-to-end vector
and the $x-y$ plane because the angle $\theta$ is approximately zero. We
can then write down the following equation:

\begin{eqnarray}
  \frac{d}{dt}\varphi=
   -\dot{\gamma}\frac{r^2-1}{r^2+1}\sin^2\varphi -
   \dot{\gamma}\frac{1}{r^2+1}
   + 2\sqrt{D_r}\zeta.
   \label{jeffrey}
\end{eqnarray}
For short times, the mean square displacement of $\varphi(t)$ in time $t$ is written as

\begin{eqnarray}
 <|{\varphi(t) - \varphi(0)}|^{2}>=4D_rt\;\;\;(for\;\;D_rt\ll1).
  \label{diffusive_angle}
\end{eqnarray}

\begin{figure}[t]
 \begin{center}
  \includegraphics[width=0.8\hsize]{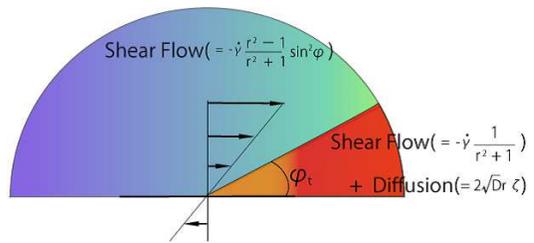}
%Fig.9
  \caption{(Color online)
  Schematic diagram of two areas and the angle $\varphi_t$. The angle
  $\varphi_t$ is the boundary angle of the area dominated by the first
  term and the area dominated by the second and third terms on the
  right-hand side of Eq. (\ref{jeffrey}).
  }
  \label{fig_phi_t}
 \end{center}
\end{figure}

In order to analyze the tumbling motion of a single chain, we only have
to consider the dynamics in the stochastic area, where the effect of
thermal fluctuation is not negligible. In the case of
$D_r\ll\dot{\gamma}$, the time required to pass through the stochastic
area is sufficiently larger than the time required to pass through the
other area. The tumbling motion can only be understood by
considering the stochastic area. The stochastic area is dominated
by the second and third terms on the right-hand side of
Eq. (\ref{jeffrey}). The remaining area is dominated by the first term
on the right-hand side of Eq. (\ref{jeffrey}). Figure \ref{fig_phi_t}
shows the two areas and the angle $\varphi_t$, which is the boundary
angle between the two areas. We call the second and third terms on the
right-hand side of Eq. (\ref{jeffrey}) the shear and the
fluctuation terms, respectively.

\subsubsection{INFINITE ASPECT RATIO CASE}

First, we consider Eq. (\ref{jeffrey}) in the limit where
$r\rightarrow\infty$. This limit is consistent with treating a chain as
an end-to-end vector. We can rewrite Eq. (\ref{jeffrey}) as
$\dot{\varphi}=-\dot{\gamma}\sin^2\varphi + 2\sqrt{D_r}\zeta$, in which
the shear term does not exist. This equation is identical to the
equation used in previous works \cite{flexible_theo1, flexible_theo2,
flexible_sim1, flexible_sim2}.

Previous works \cite{dna_exp3, flexible_sim1, flexible_sim2} have
reported $\varphi_{t}\sim(D_r/\dot{\gamma})^{1/3}$. In the region of
$\varphi_t<\varphi<\pi$, the shear flow rapidly rotates the orientation
of the chain from $\pi$ to $\varphi_{t}$ in a time $t_l \approx
\dot{\gamma}^{-1}$. In the region of $0<\varphi<\varphi_t$, the chain
orientation almost aligns along the shear direction. Because the effect
of shear becomes sufficiently small in this region, the effect of
thermal fluctuations only contributes to rotate the chain orientation
from $\varphi_{t}$ to $0$ in time
$t_r=\varphi_t^{2}/4D_r\propto\dot{\gamma}^{-2/3}$, as calculated with
Eq. (\ref{diffusive_angle}). In the case of $t_r > t_l$, $t_r$ dominates
the chain tumbling time. We can consider $\nu_p \propto
t_r^{-1}$. Therefore, we conclude from
Eqs. (\ref{def_peclet}) and (\ref{def_freq}) that $F \propto Pe^{2/3}$.

\subsubsection{FINITE ASPECT RATIO CASE}

Next, we consider Eq. (\ref{jeffrey}) with a finite aspect ratio, $r$. We
can expect $F$ to be proportional to $Pe^{2/3}$ when the fluctuation
term dominates in the region $0<\varphi<\varphi_t$, as this case
agrees with the limit of $r \rightarrow \infty$. Additionally, we can
expect $F$ to be proportional to $Pe$ when the shear term dominates in the
region $0<\varphi<\varphi_t$ because this case corresponds to the
non-Brownian rigid rod. $Pe_{\rm c}$ is defined as the Peclet number at
which the dominating term in the stochastic area changes from the
fluctuation term to the shear term with increasing shear rate.

 From Eq. (\ref{jeffrey}), we can write down the corresponding
Fokker-Planck equation as

\begin{eqnarray}
   \frac{\partial}{\partial t}P(t,\varphi) + 
    \frac{\partial}{\partial \varphi}J_{\varphi} = 0,
  \label{focker_jeffrey}
\end{eqnarray}

\begin{eqnarray}
  J_{\varphi} =
   -\left\{
     \dot{\gamma}\frac{r^2-1}{r^2+1} \sin^2 \varphi
     +\dot{\gamma}\frac{1}{r^2+1}
     +2D_r \frac{\partial }{\partial \varphi}
    \right\} P(t,\varphi),
   \label{J_phi}
\end{eqnarray}
where $P(t,\varphi)$ is the PDF of the angle $\varphi$ and $J_{\varphi}$ is
the probability flow. Each term in the braces of Eq. (\ref{J_phi})
corresponds to a respective term on the right-hand side of
Eq. (\ref{jeffrey}).

Next, we focus on the stationary PDF $P_{\rm st}(\varphi) \equiv P(t=0, \varphi)$. Because $\partial P_{\rm st}(\varphi)/ \partial t=0$, namely, $\partial J_{\varphi}/ \partial \varphi=0$ from Eq. (\ref{focker_jeffrey}), $J_{\varphi}$ is $\varphi$-independent and constant: $J_{\varphi}=-\dot{\gamma}[(r^2-1) \sin^2 \varphi_{\rm p} P_{\rm st}(\varphi_{\rm p})+P_{\rm st}(\varphi_{\rm p})]/(r^2+1)$, where $\varphi_{\rm p}$ is the angle at which $P_{\rm st}(\varphi)$ has a peak.
We surmise that the angle $\varphi_t$ satisfies the equation given by

\begin{eqnarray}
 \frac{J_{\varphi}}{2}
  = -\dot{\gamma} \frac{(r^2-1) \sin^2 \varphi P_{\rm st}(\varphi)}
  {r^2+1} \nonumber\\
 = -\dot{\gamma} \frac{P_{\rm st}(\varphi)}{r^2+1}
     -2D_r \frac{\partial P_{\rm st}(\varphi)}{\partial \varphi}.
  \label{J_phi_t_divided_2}
\end{eqnarray}
By substituting the $J_{\varphi}$ expressed by $\varphi_{\rm p}$ into
Eq. (\ref{J_phi_t_divided_2}), the relation of $\varphi_t$ to
$\varphi_{\rm p}$ is given as

\begin{eqnarray}
 -\dot{\gamma} \frac{(r^2-1) \sin^2 \varphi_{\rm p}
  P_{\rm st}(\varphi_{\rm p})+P_{\rm st}(\varphi_{\rm p})} {r^2+1}
  \notag \\ =
  -2\dot{\gamma}\frac{(r^2-1) \sin^2 \varphi_t P_{\rm st}(\varphi_t)} {r^2+1}.
  \label{J_phi_t_equal_J_phi_p}
\end{eqnarray}
The angle $\varphi_t$ is calculated by solving
Eq. (\ref{J_phi_t_equal_J_phi_p}). $P_{\rm st}(\varphi_t)$ is expanded
in powers $\triangle \varphi = \varphi_t - \varphi_{\rm p}$. When we
neglect the second order of $\triangle \varphi$ and higher, we can
relate $\varphi_t$ to $\varphi_{\rm p}$ as

\begin{eqnarray}
 \varphi_t =
  \sqrt{\frac{1}{2} \left(
		     \varphi_{\rm p}^2
		     + \frac{1}{r^2-1} \right)}.
  \label{phi_t_1}
\end{eqnarray}
It should be noted that $\varphi_t, \varphi_{\rm p}\ll1$. We can
estimate $\varphi_t$ with the angle $\varphi_{\rm p}$.

Next, we attempt to calculate the analytical form of $\varphi_{\rm p}$.
In the case of $D_r/\dot{\gamma}\ll1$, the formal solution for
$P_{\rm st}(\varphi)$ is given by

\begin{eqnarray}
  P_{\rm st}(\varphi) = C_1\int^{\pi}_0 d\psi
   \exp \left(
	 - \frac{\dot{\gamma}}{4D_r} f(\psi, \varphi)
	 \right), \label{solution_focker_jeffrey} \\
  f(\psi, \varphi) = \psi - (1-\frac{2}{r^2+1})\sin \psi \cos (\psi-2\varphi),
\end{eqnarray}
where $C_1$ is determined from the normalization condition, 
$\int^{\pi}_0 P_{\rm st}(\varphi) d\varphi = 1$.
In the limit of $r \rightarrow \infty$,
it is known that $\varphi_{\rm p}=(D_{\rm r}/\dot{\gamma})^{1/3}$
\cite{flexible_sim2}. When $r$ is finite, $\varphi_{\rm p}$ is different
from $(D_{\rm r}/\dot{\gamma})^{1/3}$ because the shear term influences
$P_{\rm st}(\varphi)$.

To estimate the effect of the shear
term, we introduce the times $t_2^{'}$ and $t_3^{'}$, where $t_2^{'}$
and $t_3^{'}$ represent the times required to pass through the region
$0<\varphi<(D_{\rm r}/\dot{\gamma})^{1/3}$, depending only on the shear
term and the fluctuation term, respectively. $t_2^{'}$ is estimated by
dividing $(D_{\rm r}/\dot{\gamma})^{1/3}$ by the shear term as

\begin{eqnarray}
  t_2^{'}=\frac{(r^2+1)D_{\rm r}^\frac{1}{3}}{\dot{\gamma}^\frac{4}{3}},
\end{eqnarray}
and $t_3^{'}$ is estimated by Eq. (\ref{diffusive_angle}) as

\begin{eqnarray}
  t_3^{'}=\frac{1}{4D_{\rm r}^\frac{1}{3}\dot{\gamma}^\frac{2}{3}}.
\end{eqnarray}

Because the Brownian chain is rotated by both the shear term and the
fluctuation terms, we expect that $\varphi_{\rm p}$ is given by 

\begin{eqnarray}
  \varphi_{\rm p}
   =C^{'}(r,Pe)\left(\frac{D_{\rm r}}{\dot{\gamma}} \right)^{\frac{1}{3}},
   \label{phi_peak}
\end{eqnarray}

\begin{eqnarray}
 \begin{array}{lll}
  C^{'}(r,Pe) &\equiv \frac{t_2^{'}}{t_2^{'}+t_3^{'}} \\
  &= \left\{ 1+\frac{1}{4r^2+1}\left(\frac{r^3}{18(\ln r + d (r))} Pe
			       \right)^{\frac{2}{3}} \right\}^{-1}.
 \end{array}
\end{eqnarray}
For large $\dot{\gamma}/D_{r}$ where the present theoretical analysis is
valid, $C^{'}(r,Pe) \sim t_{2}^{'}/t_{3}'$ finally approaches zero.
We confirmed, however, that $t_{2}^{'}/t_{3}'$ still remains finite
($\sim 1$)
around $Pe_{c} \approx 100$, where $\dot{\gamma}/D_{r}\approx 1000$.

\begin{figure}
 \begin{center}
  \includegraphics[width=1.0\hsize]{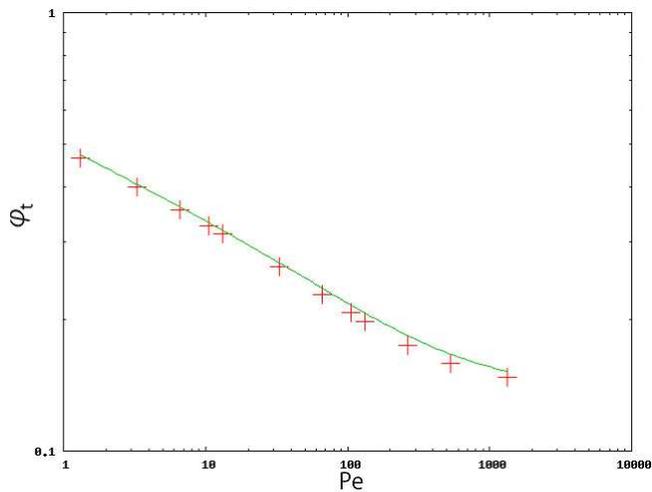}
 \end{center}
%Fig.10
 \caption{(Color online)
 The behavior of $\varphi_t$ as a function of
 $Pe$. $\varphi_{t^{'}}$ is $\varphi_{t}$ calculated with $\varphi_{\rm
 p}$, obtained from numerical integration (red $+$). The green solid line
 corresponds to the behavior of $\varphi_{t^{''}}$, where
 $\varphi_{t^{''}}$ is $\varphi_{t}$, obtained from $\varphi_{\rm p} =
 [t_2^{'}/(t_2^{'}+t_3^{'})] \times (D_{\rm r}/\dot{\gamma})^{1/3}$.
 }
 \label{graph_phi_t}
\end{figure}

Figure \ref{graph_phi_t} shows the behaviors of $\varphi_{t^{'}}$ and
$\varphi_{t^{''}}$ as a function of $Pe$, where $\varphi_{t^{'}}$ is
$\varphi_t$ calculated with Eq. (\ref{phi_t_1}) and $\varphi_{\rm p}$,
which is obtained from the numerical integration of Eq. 
(\ref{solution_focker_jeffrey}), and $\varphi_{t^{''}}$ is $\varphi_t$
calculated with Eq. (\ref{phi_t_1}) and $\varphi_{\rm p}$, which is
expressed as Eq. (\ref{phi_peak}). In this figure, it is shown that
Eq. (\ref{phi_peak}) is established because the behavior of
$\varphi_{t^{'}}$ agrees well with $\varphi_{t^{''}}$ for
$Pe\leq100$. We can obtain $\varphi_t$ by substituting
Eq. (\ref{phi_peak}) into Eq. (\ref{phi_t_1}). Therefore, $\varphi_t$ is
given by

\begin{eqnarray}
  \varphi_t =
   C(r, Pe)\left(\frac{D_{\rm r}}{\dot{\gamma}}\right)^{\frac{1}{3}},
   \label{phi_t}
\end{eqnarray}

\begin{eqnarray}
 C(r, Pe) =
  \sqrt{\frac{1}{2} \left\{
		     C^{'2}
		     + 4 \left( \frac{r^2+1}{r^2-1} \right)
		     \frac{1-C^{'}}{C^{'}}
		    \right\}}.
\end{eqnarray}

We consider that $F$ is proportional to $Pe^{2/3}$ when the effect of
thermal fluctuation is more significant than the effect of shear flow in
the region of $0<\varphi<\varphi_t$ and that $F$ is proportional to $Pe$ in
the opposite case. $t_2$ is the time required to pass through the region
$0<\varphi<\varphi_t$ by the shear term,

\begin{eqnarray}
  t_2=\frac{(r^2+1) C D_{\rm r}^\frac{1}{3}}{\dot{\gamma}^\frac{4}{3}}.
\end{eqnarray}
$t_3$ is the time required to pass through the region
$0<\varphi<\varphi_t$ by the fluctuation term,
\begin{eqnarray}
  t_3=\frac{C^2}{4D_{\rm r}^\frac{1}{3}\dot{\gamma}^\frac{2}{3}}.
\end{eqnarray}

It is thought that $Pe_{\rm c}$ is the Peclet number that satisfies
$t_2=t_3$. If we obtain the value of $r$, we know the value of $Pe_{\rm
c}$ because both $t_2$ and $t_3$ are functions of $Pe$ and $r$. Our
results show that $Pe_{\rm c}\approx115$ at $r=5$. In the numerical
results obtained from our work, $Pe_{\rm c}=106$. The analytical result
agrees well with our numerical result. Moreover, the numerical condition
$D_{\rm r}\ll\dot{\gamma}$ is satisfied because $D_{\rm
r}/\dot{\gamma}\approx10^{-3}\ll1$ at $Pe\approx100$. Therefore, the
considerations in this section are reasonable in the region near
$Pe_{\rm c}$.

The considerations in this section agree well with those of previous experimental
results \cite{chu_science, dna_exp1, dna_exp2, dna_exp3,
free_rotation}. In experimental works that measured the frequencies of
DNA rotation \cite{chu_science, dna_exp1, dna_exp2, dna_exp3}, the DNA
molecules contained roughly 400 persistence lengths. The persistence length is thought to correspond to $r$. Thus, we conclude that
$Pe_{\rm c}\approx516$. These experiments were carried out in the region
of $Pe \ll Pe_{\rm c}$, and $F$ is proportional to $Pe^{2/3}$. In
experimental work that measured the frequencies of freely rotating rigid
dumbbells \cite{free_rotation},the aspect ratio $r$ of the rigid
dumbbell corresponds to $2$ and $Pe_{\rm c}\approx60$. The experiments
were carried out in the region of $Pe_{\rm c} \ll Pe$, and $F$ was
proportional to $Pe$. From these results, the considerations in this
section are reasonable.

\section{CONCLUSION}

In our work, we calculated the tumbling motion of a single chain using
an SPM that takes into account thermal fluctuations and hydrodynamic
interactions for $0.0005<k_{\rm B}T<1.00$ and
$0.001<\dot{\gamma}<0.02$. We conclude that the dimensionless frequency,
$F$, depends only on $Pe$. The dependence of $F$ can be described by a
power law $F \propto Pe^{\alpha}$. The exponent $\alpha$ sharply changes
from $2/3$ to $1$ on $Pe_{\rm c}$. In the case of a rigid rod with
$N=5$, $Pe_{\rm c}\approx106$, and in the case of a flexible chain with
$N=5$, $Pe_{\rm c}\approx156$. The behavior of $F$ for both cases is
similar, while only the values of $Pe_{\rm c}$ are different from each
other.

We have presented $F$ to be proportional to $Pe^{2/3}$ when the third
term on the right-hand side of Eq. (\ref{jeffrey}) dominates in the
region $0<\varphi<\varphi_t$, and $F$ is proportional to $Pe$ when the
second term of Eq. (\ref{jeffrey}) dominates in the region
$0<\varphi<\varphi_t$. We have estimated the angle $\varphi_t$ at which
the fist term of $J_{\varphi}$, expressed as Eq. (\ref{J_phi}), is
comparable to the sum of the second and third terms of $J_{\varphi}$,
expressed as Eq. (\ref{J_phi}).

A proposed mechanism for this exponent change is that the effect of
thermal fluctuation is more significant than the effect of shear flow
only for $0<\varphi<\varphi_t$, whereas in the other case, the effect of
thermal fluctuation is negligible. The former contribution leads to $F
\propto Pe^{2/3}$, and the latter contribution leads to $F \propto Pe$.

\section{ACKNOWLEDGEMENT}

The authors would like to express their gratitude to Dr. T. Murashima,
Dr. Y. Nakayama, Dr. K .Kim, and Dr. T .Iwashita for useful comments and
discussions.

\end{document}